\documentclass[11pt,a4paper,twosides]{amsart}

\usepackage{stmaryrd}
\usepackage{amsmath}
\usepackage{amssymb}
\usepackage{cite}
\usepackage{tabularx}
\usepackage[top=2.54cm,bottom=3.54cm,left=3.15cm,right=3.15cm ]{geometry}
\usepackage[bookmarksnumbered=true]{hyperref}

\input psfig.sty

\newcommand{\dd}{\mathrm{d}}
\newcommand{\bR}{\mathbb{R}}
\newcommand{\pt}{\partial_t}
\newcommand{\px}{\partial_x}
\newcommand{\pxx}{\partial_{xx}}

\newcommand{\fexp}{~\mathrm{exp}}
\newcommand{\sech}{~\mathrm{sech}}

\begin{document}


\title[Computation for coupled Klein--Gordon]
{Numerical solutions of the coupled nonlinear Klein--Gordon equations by trigonometric integrator pseudospectral discretization}

\date{\today}

\author[X. Dong]{Xuanchun Dong}

 \address{{\bf Xuanchun Dong}\newline
 Center for Computational Science and Engineering\newline
          Department of Mathematics\\
          National  University of Singapore\newline
          Block S17, 10, Lower Kent Ridge Road, 119076, Singapore}
 \email{dong.xuanchun@nus.edu.sg\\
        dong.xuanchun@gmail.com}


\maketitle

\begin{abstract}
A scheme stemming from the use of pseudospectral approximations to
spatial derivatives followed by a time integrator based on trigonometric polynomials is proposed for the numerical solutions of the coupled nonlinear Klein--Gordon equations.  Numerical tests on one- and
three-coupled Klein--Gordon equations are presented, which are geared
towards understanding the accuracy and stability, and demonstrating
the efficiency and high resolution capacity in application.
\end{abstract}

\section{Introduction}
\setcounter{equation}{0}

The characteristics of nonlinear phenomena in various physics fields
as, e.g., fluid dynamics, laser and fiber optics, solid state
physics,  plasma physics, chemical physics, reaction kinematics and
etc., can be mathematically described by nonlinear evolution
equations \cite{AC_book}.  Among them of great physical significance
are the equations that possess soliton solutions.  In recent decades
a variety of methods, such as the inverse scattering method
\cite{AC_book,AS_SIAM}, bilinear transformation
\cite{Hirota_bilinear} and etc., have been developed for obtaining
the exact solutions of soliton type equations.  In parallel with the
analytical treatment a surge of studies have been devoted to the
numerics of these equations, which is a topic of great importance in
applied science. In the present work the numerics of coupled
nonlinear Klein--Gordon equations governing waves in a dispersive media
is going to be examined.

$N$-coupled nonlinear Klein--Gordon equations under consider take the
general form \cite{ACN_ckg,HO_ckg}:
\begin{align}
  & \left(\frac{\partial^2}{\partial t^2}-\frac{\partial^2}{\partial x^2}\right)
  \psi_k + \psi_k - 2\left(\sum_{p=1}^N\psi_p^2+Q\right)\psi_k =
  0,\label{ckg1}\\
  & \left(\frac{\partial}{\partial t}-\frac{\partial}{\partial
  x}\right)Q + 2\frac{\partial}{\partial t}\sum_{p=1}^N\psi_p^2=0,\quad x\in\bR,\quad
  t>0,\label{ckg2}
\end{align}
with $k=1,2,\ldots,N$, for $\psi_k(x,t)$ and $Q(x,t)$ sufficiently
differentiable functions. Note that there exists an invariant of
(\ref{ckg1})-(\ref{ckg2}):
\begin{equation}\label{energy}
 E(t) = \int_{\bR} \bigg\{\sum_{k=1}^N\left[\left(\pt \psi_k\right)^2 + \left(\px \psi_k\right)^2
  +\psi_k^2\right]-\bigg(\sum_{k=1}^N\psi_k^2\bigg)^2
  +\frac{1}{2}Q^2\bigg\}\dd x,\quad t\geq 0,
\end{equation}
and it suffices to refer the above conserved quantity as {\it
energy}.  Such conservation law can be justified from multiplying
both sides of (\ref{ckg1}) by $2 \pt\psi_k$ and, respectively, both
sides of (\ref{ckg2}) by $Q$, integrating over $\bR$ and then
summing up for $k=1,2,\ldots,N$. Bilinear form and one-soliton
solutions of (\ref{ckg1})-(\ref{ckg2}) were investigated in
\cite{ACN_ckg,HO_ckg}, and the complete integrability was also
constructed from P-analysis \cite{ACN_integrable,PA_P-analysis}.
Along the numerical aspects of (\ref{ckg1})-(\ref{ckg2}), to our
best knowledge there are few results derived in literature.

The goal of this paper is to propose an efficient and accurate
numerical scheme for solving (\ref{ckg1})-(\ref{ckg2}). The key
point in designing the scheme is based on applying Fourier
pseudospectral approximations to the spatial derivatives, followed
by applying a time integrator based on trigonometric polynomials, i.e.,
the so-called Gautschi-type or Deuflhard-type integrator (see e.g.
\cite{Deu79,Gau61}) in phase space to the temporal discretization.
The resulting scheme is fully explicit, symmetric in time,
spectral-order of accuracy in space and second-order of accuracy in
time, easy to implement and with less memory demand. Moreover,
numerical tests demonstrate that the scheme is stable and conserves
the energy defined by (\ref{energy}) very well, which are two
favorable properties desired for a scheme in long-time simulation
application. Note that the similar technique has been used for
standard non-coupled Klein--Gordon equations \cite{BD2011}, and for
other classes of schemes, as for instance the finite-difference,
decomposition and etc., for non-coupled Klein--Gordon equations we
refer the readers to \cite{CaoGuo,Deeba,Duncan,LiVu,Pascual} and
references given therein.

The rest is as follows. In Section \ref{s2} efficient numerical
scheme is proposed. In Section \ref{s3} numerical results are
reported for accuracy and stability tests of the scheme, and its
application in numerically studying the dynamics of soliton-soliton
collisions in one- and three-coupled Klein--Gordon equations. Finally
some conclusions are drawn in Section \ref{s4}.

\section{Numerical scheme}\label{s2}
\setcounter{equation}{0}

In this section we shall propose the efficient numerical
discretization. The initial conditions are assumed to be of the
form:
\begin{equation}\label{initial_cond}
   \psi_k(x,t=0) = \psi_k^{(0)}(x),\quad
  \pt\psi_k(k,t=0)=\psi_k^{(1)}(x),\quad
  Q(x,t=0)=Q^{(0)}(x),\quad x\in \bR.
\end{equation}
In practice we truncate the whole space problem into an interval
$[a,b]$ with periodic boundary conditions
\begin{equation*}
  \psi_k(a,t)=\psi_k(b,t),\quad\px\psi_k(a,t)=\px\psi_k(b,t),\quad
  Q(a,t)=Q(b,t),\qquad t\geq 0.
\end{equation*}

Choose mesh size $h:=(b-a)/M$ with $M$ an even positive integer,
time step $\tau>0$, and denote the grid points with coordinates
$(x_j,t_n)=(a+jh,n\tau)$ for $j=0,1,\ldots,M$ and $n=0,1,\ldots$.
Let $(\psi_k)_j^n$ and $Q_j^n$ be the approximations of
$\psi_k(x_j,t_n)$ and $Q(x_j,t_n)$, and denote by $\psi_k^n$ and
$Q^n$ the solution vectors with components$(\psi_k)_j^n$ and $Q_j^n$
respectively.

\subsection{Discretization for (\ref{ckg1})}

Assume
\begin{equation}\label{app_psi}
  \psi_k(x,t)\approx
  \sum_{l=-M/2}^{M/2-1}\widetilde{(\psi_k)}_l(t)\;\fexp\left(i\mu_l(x-a)\right),
\end{equation}
for $ a\leq x \leq b$ and $t\geq 0$, with $\mu_l=2\pi l/(b-a)$
($l=-M/2,\ldots,M/2-1$) and $\widetilde{(\psi_k)}_l(t)$ the discrete
Fourier transformation coefficient of the $l$th mode,
\begin{equation}
  \widetilde{(\psi_k)}_l(t) = \frac{1}{M}\sum_{j=0}^{M-1}
  \psi_k(x_j,t)\fexp\left(-i\mu_l(x_j-a)\right),\quad l=-M/2,\ldots,M/2-1.
\end{equation}
Approximating the spatial derivatives in (\ref{ckg1}) by Fourier
pseudospectral discretization \cite{ST}, i.e.,
\begin{equation*}
  -\pxx \psi_k(x,t)\approx
  \sum_{l=-M/2}^{M/2-1}\mu_l^2\widetilde{(\psi_k)}_l(t)\;\fexp\left(i\mu_l(x-a)\right),
\end{equation*}
and noting orthogonality of the Fourier functions, we obtain the
following second-order ODEs in phase space, for
$l=-M/2,\ldots,M/2-1$,
\begin{equation}
  \frac{\dd^2}{\dd t^2}\widetilde{(\psi_k)}_l(t)
  + \left(\mu_l^2+1\right)\widetilde{(\psi_k)}_l(t) -
  \widetilde{\left(f_k\right)}_l(t) = 0,
\end{equation}
with
$f_k(x,t)=2\left(\sum_{p=1}^N\psi_p^2(x,t)+Q(x,t)\right)\psi_k(x,t)$
and $\widetilde{\left(f_k\right)}_l(t)$ defined in an analogous way
as (\ref{app_psi}). The analytical solutions of the above
second-order ODEs can be written explicitly thanks to
variation-of-constants formula. For $t_n$ (n=0,1,\ldots) a given
time,
\begin{align}
  \widetilde{(\psi_k)}_l(t) = &~
  \widetilde{(\psi_k)}_l(t_n)\cos\left(\lambda_l(t-t_n)\right) +\lambda_l^{-1}\frac{\dd}{\dd t}\widetilde{(\psi_k)}_l(t_n)\sin\left(\lambda_l(t-t_n)\right)\nonumber\\
  &+\lambda_l^{-1}\int_{t_n}^{t}
  \widetilde{\left(f_k\right)}_l(s)\sin\left(\lambda_l(t-s)\right)\dd
  s,\label{sol_ODEs}
\end{align}
with $\lambda_l=\sqrt{\mu_l^2+1}$.

The approximations of $\psi_k(x,t_1)$ are achieved from
(\ref{sol_ODEs}) with $n=0$ and $t=t_1=\tau$ together with the
initial conditions (\ref{initial_cond}) and approximating the
integrals by standard trapezoidal rule.
For $n=1,2,\ldots$, adding (\ref{sol_ODEs}) with
$t=t_{n+1}=t_n+\tau$ to its evaluation at $t=t_{n-1}=t_n-\tau$, and
then approximating the integrals via trapezoidal rule, we get the
three-term recurrence:
\begin{align}
  \widetilde{(\psi_k)}_l(t_{n+1}) =&-\widetilde{(\psi_k)}_l(t_{n-1})
  +  2\widetilde{(\psi_k)}_l(t_n)\cos\left(\lambda_l\tau\right)\nonumber\\
  &+\lambda_l^{-1}\int_{0}^{\tau}\left[
  \widetilde{\left(f_k\right)}_l(t_n+s)+\widetilde{\left(f_k\right)}_l(t_n-s)\right]
  \sin\left(\lambda_l(\tau-s)\right)\dd
  s\nonumber\\
  \approx &-\widetilde{(\psi_k)}_l(t_{n-1})
  +  2\widetilde{(\psi_k)}_l(t_n)\cos\left(\lambda_l\tau\right)
   +\tau\lambda_l^{-1}\widetilde{\left(f_k\right)}_l(t_n)\sin(\lambda_l\tau).
\end{align}


\subsection{Discretization for (\ref{ckg2})}

 Again,
assuming
\begin{equation*}
  Q(x,t)\approx \sum_{l=-M/2}^{M/2-1}\widetilde{Q}_l(t)~\fexp\left(i\mu_l(x-a)\right),
\end{equation*}
for $a\leq x\leq b,\;t\geq 0$, and approximating the spatial
derivative in (\ref{ckg2}) by Fourier pseudospectral discretization
\cite{ST},
\begin{equation}
  \px Q(x,t)\approx \sum_{l=-M/2}^{M/2-1} i\mu_l\widetilde{Q}_l(t)~\fexp\left(i\mu_l(x-a)\right),
\end{equation}
we have the following first-order ODEs in phase space, for
$l=-M/2,\ldots,M/2-1$,
\begin{equation}\label{ODE_Q}
  \frac{\dd}{\dd t}\widetilde{Q}_l(t) - i\mu_l \widetilde{Q}_l(t) =
  \frac{\dd}{\dd t}\widetilde{g}_l(t),
\end{equation}
 with
$g(x,t)=-2\sum_{p=1}^N\psi_p^2(x,t)$.  For $t_n$ ($n=0,1,\ldots$) a
given time, integrating (\ref{ODE_Q}) from $t=t_n$ to
$t=t_{n+1}=t_n+\tau$ and approximating the integrals via trapezoidal
rule, we get,
\begin{align}
  \widetilde{Q}_l(t_{n+1})=&~\fexp\left(i\mu_l\tau\right)\widetilde{Q}_l(t_n)
  +\widetilde{g}_l(t_{n+1})-\fexp\left(i\mu_l\tau\right)\widetilde{g}_l(t_{n})\nonumber\\
 & +i\mu_l
  \int_{t_n}^{t_{n+1}}\fexp\left(i\mu_l(t_{n+1}-t)\right)\widetilde{g}_l(t)\dd
  t\nonumber\\
  \approx & ~\fexp\left(i\mu_l\tau\right)\widetilde{Q}_l(t_n) +
  \left(\frac{i\mu_l\tau}{2}+1\right)\widetilde{g}_l(t_{n+1})
  +\left(\frac{i\mu_l\tau}{2}-1\right)\fexp\left(i\mu_l\tau\right)\widetilde{g}_l(t_n).
\end{align}

\subsection{Detailed numerical scheme}

Choosing $(\psi_k)^0_j=\psi_k^{(0)}(x_j)$ and $Q_j^0=Q^{(0)}(x_j)$,
the detailed numerical scheme is as follows: for $n=1,2,\ldots$ and
$j=0,1,\ldots,M$
\begin{align}
  &(\psi_k)_j^n =
  \sum_{l=-M/2}^{M/2-1}\widetilde{\left(\psi_k^n\right)}_l~\fexp\left(\frac{2ijl\pi}{M}\right),\label{scheme_1}\\
  & Q_j^n =
  \sum_{l=-M/2}^{M/2-1}\widetilde{\left(Q^n\right)}_l~\fexp\left(\frac{2ijl\pi}{M}\right),
\end{align}
where,
\begin{align}
  &\widetilde{\left(\psi_k^1\right)}_l =
  \widetilde{\left(\psi_k^{(0)}\right)}_l\cos\left(\lambda_l\tau\right)+
  \widetilde{\left(\psi_k^{(1)}\right)}_l\lambda_l^{-1}\sin\left(\lambda_l\tau\right)
  +\frac{\tau}{2}\lambda_l^{-1}\widetilde{\left(f^0_k\right)}_l\sin\left(\lambda_l\tau\right),\\
  \displaybreak[1]
  & \widetilde{\left(\psi_k^{n+1}\right)}_l = -\widetilde{(\psi_k^{n-1})}_l
  +  2\widetilde{(\psi_k^n)}_l\cos\left(\lambda_l\tau\right)
  +\tau\lambda_l^{-1}\widetilde{\left(f_k^n\right)}_l\sin(\lambda_l\tau),\quad
  n=1,2,\ldots,\\
  \displaybreak[1]
  & \widetilde{\left(Q^{n+1}\right)}_l = \fexp\left(i\mu_l\tau\right)\widetilde{\left(Q^n\right)}_l
   + \left(\frac{i\mu_l\tau}{2}+1\right)\widetilde{\left(g^{n+1}\right)}_l  \nonumber\\
   &\qquad\qquad+\left(\frac{i\mu_l\tau}{2}-1\right)\fexp\left(i\mu_l\tau\right)\widetilde{\left(g^n\right)}_l,\quad
   n=0,1,\ldots.
\end{align}
Here, the vector $f_k^n=\left[\left(f_k\right)_0^n,
\left(f_k\right)_1^n,\ldots,\left(f_k\right)_M^n\right]^T$ and
$g^n=\left[g^n_0,g^n_1,\ldots,g^n_M\right]^T$ are defined by
\begin{align*}
  &\left(f_k\right)_j^n = 2\left(\sum_{p=1}^N
  \left(\left(\psi_p\right)^n_j\right)^2+Q_j^n\right)\left(\psi_k\right)^n_j,\qquad
  g^n_j = -2\sum_{p=1}^N
  \left(\left(\psi_p\right)^n_j\right)^2.
\end{align*}

 If the energy defined by (\ref{energy}) is of
interests, then the approximation of $\pt\psi_k(x_j,t_n)$,
\begin{equation}
 (\phi_k)_j^n =
  \sum_{l=-M/2}^{M/2-1}\widetilde{\left(\phi_k^n\right)}_l~\fexp\left(\frac{2ijl\pi}{M}\right),
\end{equation} for $j=0,1,\ldots,M$ and $n=1,2,\ldots$ can be obtained
from
\begin{equation}\label{scheme_last}
  \widetilde{\left(\psi_k^{n+1}\right)}_l-\widetilde{\left(\psi_k^{n-1}\right)}_l
  =2\lambda_l^{-1}\widetilde{\left(\phi_k^n\right)}_l\sin\left(\lambda_l\tau\right),
\end{equation}
which is achieved by subtracting (\ref{sol_ODEs}) with
$t=t_{n+1}=t_n+\tau$ from its evaluation at $t=t_{n-1}=t_n-\tau$,
and applying trapezoidal rule to the integrals.

The scheme (\ref{scheme_1})-(\ref{scheme_last}) is fully explicit,
symmetric in time by noting that it is unchanged if we interchange
$\tau\leftrightarrow-\tau$ and $n\leftrightarrow n+1$, and quite
efficient in implementation thanks to FFT.  It is of spectral-order
of accuracy in space, i.e., it converges exponentially fast as mesh
size refined smaller, which is an expected property for the
spectral-type discretization. Moreover, as shown by numerical
experiments reported in the next section, the scheme is of
second-order of accuracy in time, stable and conserves the energy
defined by (\ref{energy}) very well.

\section{Numerical results}\label{s3}
\setcounter{equation}{0}

Numerical examples are reported in this section towards
understanding the accuracy and stability of the numerical scheme
(\ref{scheme_1})-(\ref{scheme_last}), and applying it to study
soliton-soliton collisions in one- and three-coupled nonlinear
Klein--Gordon equations.


\begin{table}[t!]
  \caption{Accuracy tests results: (i) upper part for discretization error in space, under $\tau=0.0001$;
  (ii) middle part for discretization error in time, under $h=1/8$ and
  (iii) lower part for conserved quantity analysis, under $h=1/4$ and $\tau=0.02$.}\label{tab_accuracy}
\begin{center}
\def\temptablewidth{0.8\textwidth}
\begin{tabular*}{\temptablewidth}{@{\extracolsep{\fill}}lllll}\hline
 & $h=1/2$ & $h=1/4$ & $h=1/8$ & $h=1/16$ \\[0.25em]
$e(60)$  & 1.1677E-1 & 2.8638E-6 & 1.7098E-6& 1.0244E-8\\[0.25em]
\hline
 & $\tau=0.04$ & $\tau=0.02$ & $\tau=0.01$ & $\tau=0.005$\\[0.25em]
 $e(60)$ & 1.1654E-1 & 2.9405E-2 & 7.3659E-3 & 1.8423E-3\\[0.25em]
\hline $t$ & $t=50$ & $t=100$ & $t=150$ & $t=200$\\[0.25em]
$E(t)$ & 0.67890052 & 0.67890052 & 0.67890050 & 0.67890051\\[0.25em]\hline
\end{tabular*}
\end{center}
\end{table}


\subsection{Accuracy tests and stability study}
It is known that the $N$-coupled nonlinear Klein--Gordon equations
admit the following analytical one-soliton solutions \cite{ACN_ckg}:
\begin{align}
  &\psi_{(c,\;\alpha_k)}(x,t)= \alpha_k\sqrt{\frac{1+c}{1-c}}\sech\left(\frac{x-ct}{\sqrt{1-c^2}}\right),\\
  &Q_c(x,t)=\frac{2c}{c-1}\sech^2\left(\frac{x-ct}{\sqrt{1-c^2}}\right),
\end{align}
with $\left|c\right|<1$ an arbitrary constant, and coefficients
$\alpha_k$ satisfying $\sum_{k=1}^{N}\alpha_k^2=1$.  To test the
accuracy, we solve the one-coupled nonlinear Klein--Gordon equations
(i.e. $N=1$ in (\ref{ckg1})-(\ref{ckg2})) on $[-24,104]$ for $0\leq
t\leq 200$, with initial data
$\psi_1^{(0)}(x)=\psi_{(0.4,\;1)}(x,t=0)$,
$\psi_1^{(1)}(x)=\pt\psi_{(0.4,\;1)}(x,t=0)$ and
$Q^{(0)}(x)=Q_{0.4}(x,t=0)$. To quantify the numerical results, the
error function $e(t)$ is defined as the maximum error,
$e(t_n):=\max_{j}\left|(\psi_1)^n_j-\psi_1(x_j,t_n)\right|
+\max_{j}\left|Q^n_j-Q(x_j,t_n)\right|$.


\begin{figure}[t!]
  \centerline{\; \psfig{figure=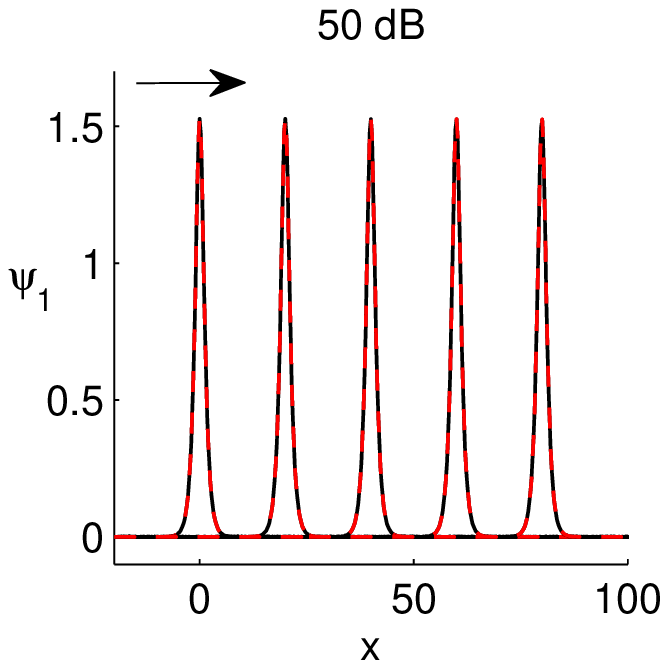,width=5cm}\psfig{figure=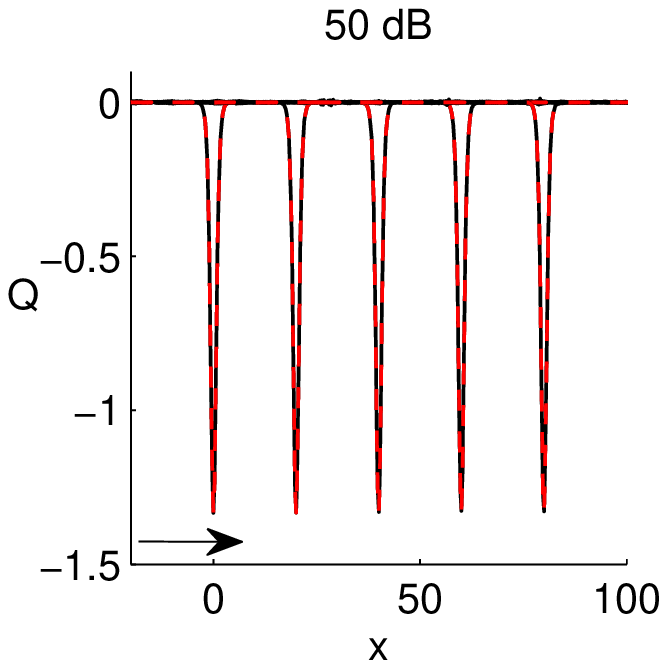,width=5cm}}
  \caption{Comparisons between numerical one-soliton results (solid line) from initial data
  perturbed by adding white Gaussian noise with signal-to-noise ratio 50dB
  and results (dashed line and red colored online) from initial data without perturbation, under $h=1/4$ and $\tau=0.02$.
  Wave fronts depicted for $t=0,~50,~100,~150$ and $200$ (from left to
  right).}\label{fig_stab}
\end{figure}

To test the discretization error in space, we choose a very fine
time step $\tau=0.0001$ such that the error from time discretization
is negligible compared to the spatial discretization error.
Similarly, we choose a very fine mesh size $h=1/8$ to eliminate the
spatial discretization error for testing the discretization error in
time. Also, mesh size $h=1/4$ and time step $\tau=0.02$ are chosen
to test the conservation of conserved quantity (\ref{energy}). The
results are listed in Table \ref{tab_accuracy}.  To study the
stability of the scheme, the same initial conditions as chosen
previously are used, to which we add white Gaussian noise with
signal-to-noise ratio 50dB. The results are depicted in Figure
\ref{fig_stab}.

Based on these results, the following are drawn:
\begin{enumerate}
  \item The numerical scheme (\ref{scheme_1})-(\ref{scheme_last}) is of spectral-order of accuracy in space,
  i.e., its convergence order in space is higher than any
  polynomial order, and of second-order of accuracy in time.
  \item It conserves the invariant defined by
  (\ref{energy}) very well (up to seven significant digits in the reported example) over a long-time simulation.
  \item It is numerically stable, by which we mean a minor
  perturbation in the initial data dose not bring in significant
  difference to the results over a long-time simulation and no numerical blow-up occurs.
\end{enumerate}

\subsection{Applications on soliton-soliton collisions}


\begin{figure}[t!]
  \centerline{\psfig{figure=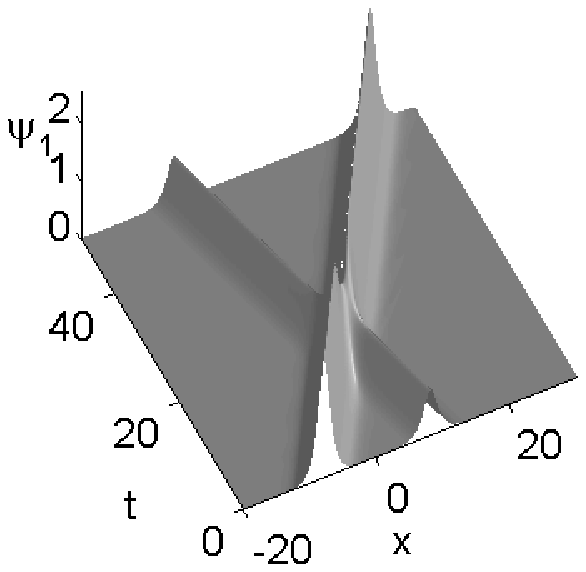,width=5cm}\psfig{figure=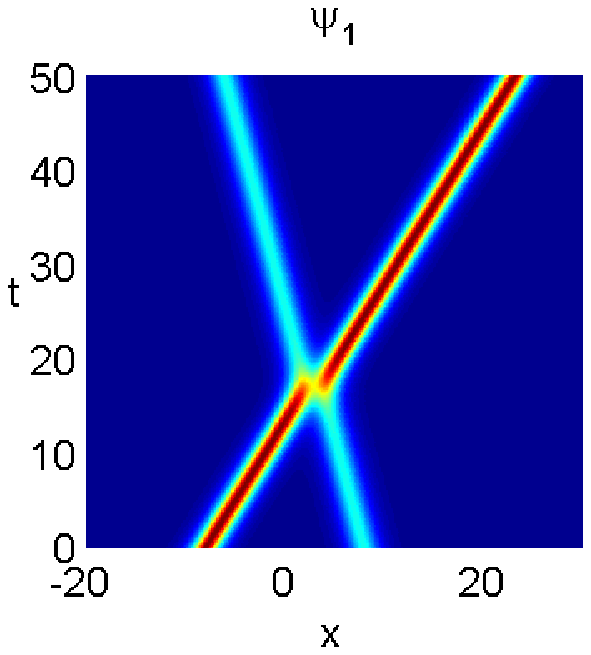,width=5cm}}
  \centerline{\psfig{figure=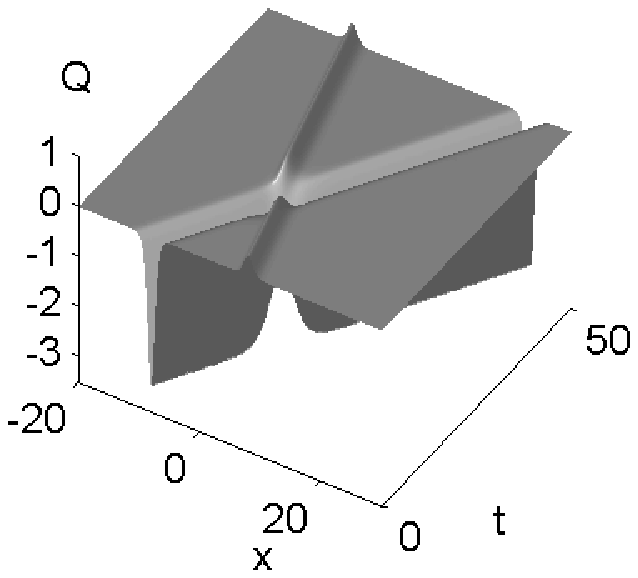,width=5cm}\psfig{figure=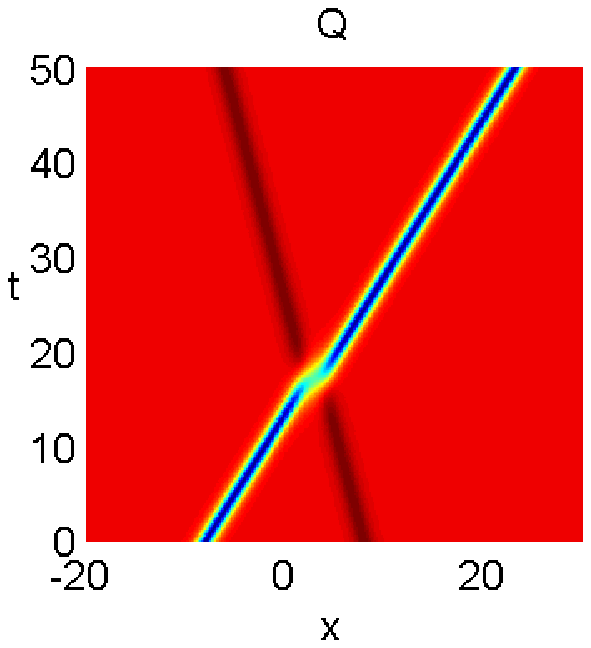,width=5cm}}
  \caption{Numerical results for soliton-soliton collisions in one-coupled Klein--Gordon equations.
  Left column for solution evolution and right column for its contours (color online).
  Computations are  carried out on $[-24,40]$ under $h=1/4$ and $\tau=0.02$.
  Initial conditions are chosen as (\ref{1sc_initial_cond_psi})-(\ref{1sc_initial_cond_Q}) with
  $x_0=8$ such that initially two one-soliton waves are well
  separated.}\label{fig_1sc1}
\end{figure}

\begin{figure}[t!]
  \centerline{\psfig{figure=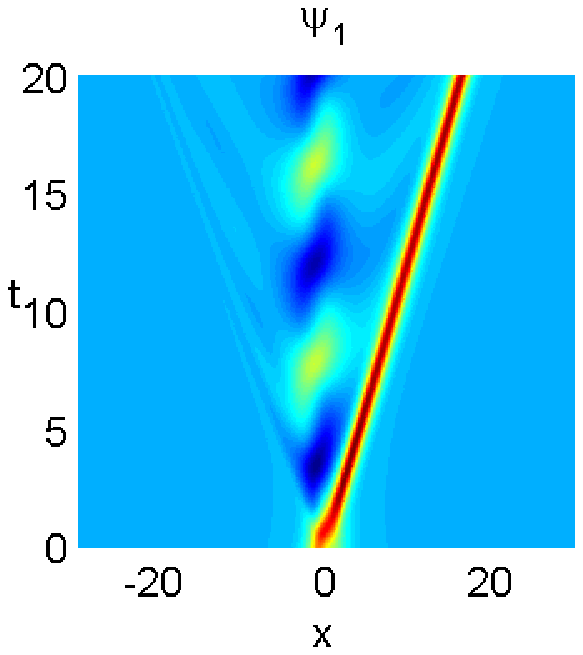,width=5cm}
  \psfig{figure=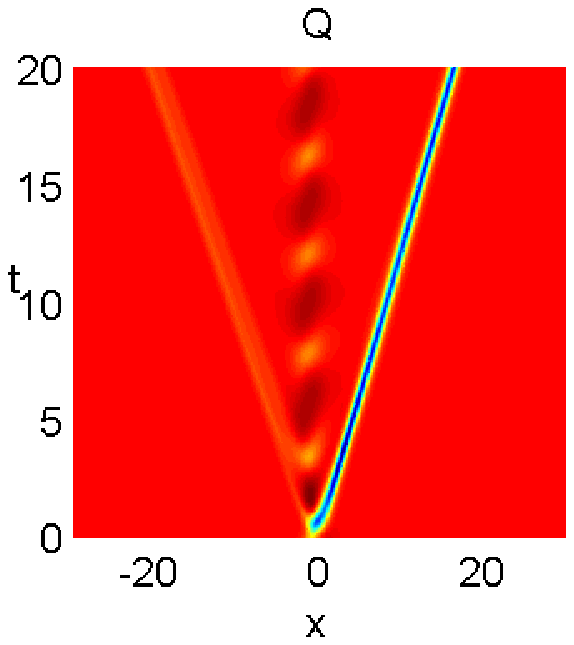,width=5cm}}
  \caption{Contours (color online) of numerical results for soliton-soliton collisions in one-coupled Klein--Gordon equations.
  Computations are  carried out on $[-32,32]$ under $h=1/4$ and $\tau=0.02$.
  Initial conditions are chosen as (\ref{1sc_initial_cond_psi})-(\ref{1sc_initial_cond_Q}) with
  $x_0=1$ such that initially two one-soliton waves are almost at the same
  location.}\label{fig_1sc2}
\end{figure}


First we apply the method to one-coupled nonlinear Klein--Gordon
equations, i.e., $N=1$ in (\ref{ckg1})-(\ref{ckg2}), with initial
date:
\begin{align}
  &\psi_1^{(0)}(x) = \psi_{(0.6,\;1)}(x+x_0,t = 0)
                    +\psi_{(-0.25,\;1)}(x-x_0,t = 0),\label{1sc_initial_cond_psi}\\
  &\psi_1^{(1)}(x) = \pt\psi_{(0.6,\;1)}(x+x_0,t = 0)
                    +\pt\psi_{(-0.25,\;1)}(x-x_0,t =
                    0),\label{1sc_initial_cond_psi_t}\\\displaybreak[1]
  &Q^{(0)}(x) =   Q_{0.6}(x+x_0,t=0)+Q_{-0.25}(x-x_0,t = 0),\label{1sc_initial_cond_Q}
\end{align}
with a dislocation parameter $x_0>0$ to measure the separation of
two one-soliton waves at initial time level.  Some results from
numerical experiments are shown in Figure \ref{fig_1sc1} and Figure
\ref{fig_1sc2}. These results indicate that when two solitons are of
nearly completed separation at initial time level, after the
collision the solutions remain to propagate in soliton pattern as
the time proceeds and there is no visible wave structure emitted. On
the other hand, when two solitons are not quite separated at initial
time level, after the collision the emission of new waves is
conspicuous.

\begin{figure}[t!]
  \centerline{\psfig{figure=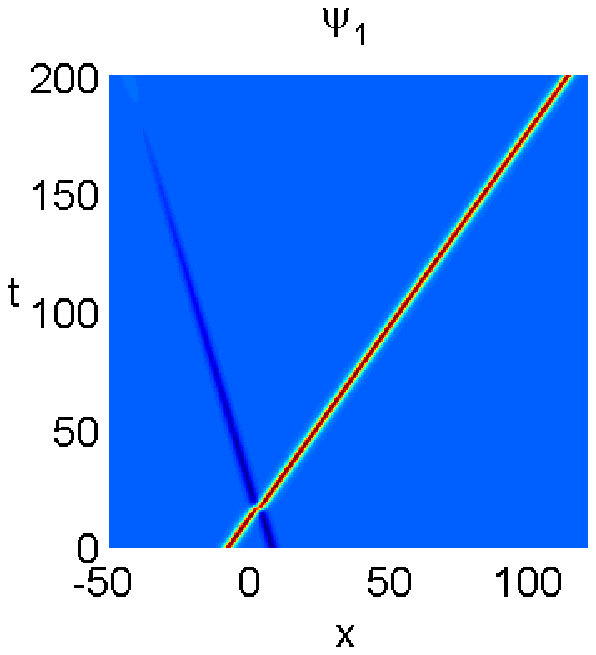,width=5cm}
  \psfig{figure=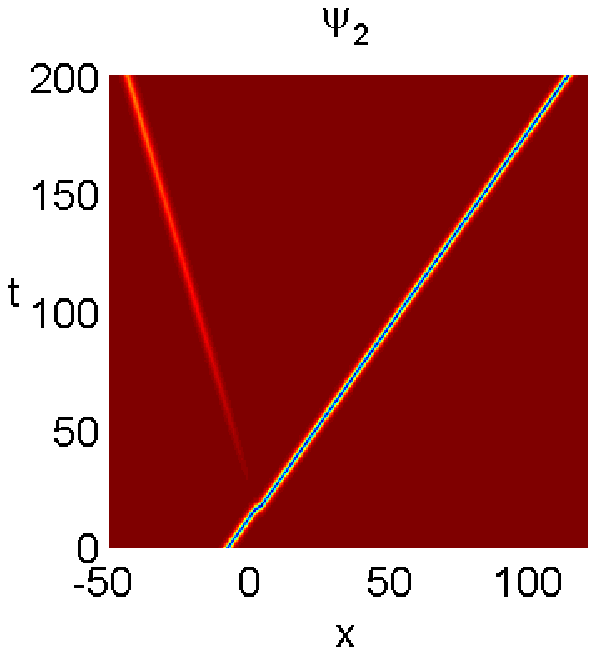,width=5cm}}
  \centerline{\psfig{figure=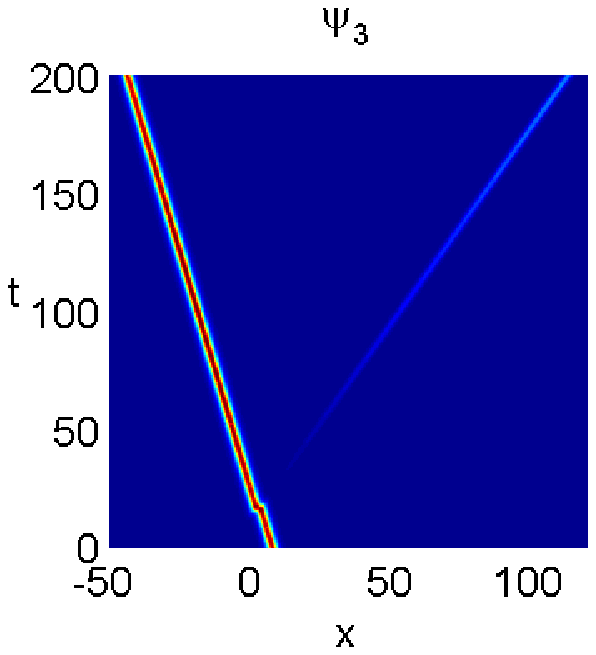,width=5cm}
  \psfig{figure=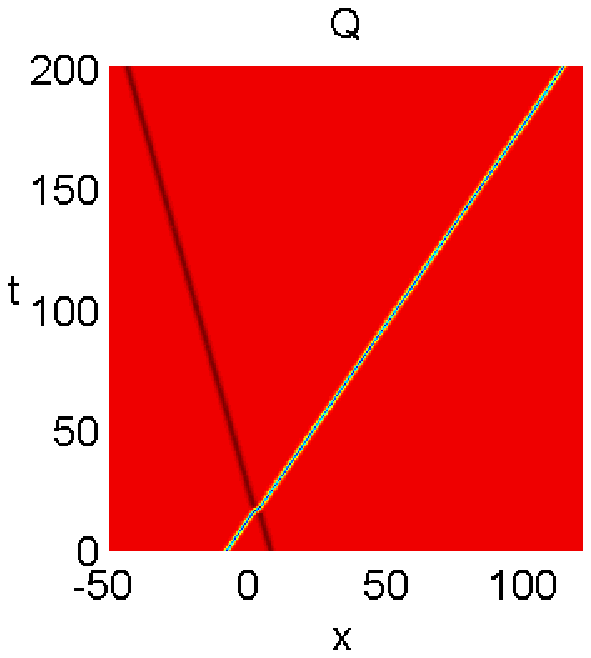,width=5cm}}
    \caption{Contours (color online) of numerical results for soliton-soliton collisions in three-coupled Klein--Gordon equations.
  Computations are  carried out on $[-96,160]$ under $h=1/4$ and $\tau=0.02$.
  Initial conditions are chosen as (\ref{3sc_initial_psi})-(\ref{3sc_initial_q}) with
  $x_0=8$ such that initially two one-soliton waves are well
  separated.}\label{fig_3sc_c}
\end{figure}

\begin{figure*}[t!]
  \centerline{\psfig{figure=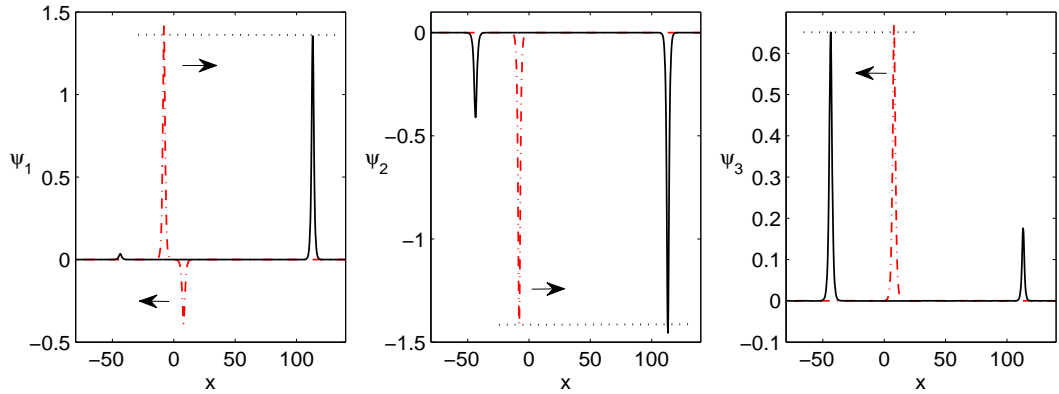,width=16.8cm}}
  \caption{Numerical results of $\psi_1$, $\psi_2$ and $\psi_3$ for soliton-soliton collisions in three-coupled Klein--Gordon equations:
  solid line corresponds to time $t=200$ and dash-dot line (red colored online) corresponds to time $t=0$.
  Parameters are same as Figure
  \ref{fig_3sc_c}.}\label{fig_3sc_evol}
\end{figure*}

Next we present an example of applying the method to three-coupled
nonlinear Klein--Gordon equations, i.e., $N=3$ in
(\ref{ckg1})-(\ref{ckg2}), and the initial conditions are taken as
\begin{align}
   & \psi_1^{(0)}(x) =  \psi_{(0.6,\;1/\sqrt{2})}(x+x_0,t=0)
    + \psi_{(-0.25,\;-1/2)}(x-x_0,t=0),\label{3sc_initial_psi}\\\displaybreak[1]
   &\psi_2^{(0)}(x) =  \psi_{(0.6,\;-1/\sqrt{2})}(x+x_0,t=0),\\\displaybreak[1]
   &\psi_3^{(0)}(x)=  \psi_{(-0.25,\;\sqrt{3}/2)}(x-x_0,t=0),\\
   &\psi_1^{(1)}(x) =  \pt\psi_{(0.6,\;1/\sqrt{2})}(x+x_0,t=0)
   +  \pt\psi_{(-0.25,\;-1/2)}(x-x_0,t=0),\\\displaybreak[1]
   &\psi_2^{(1)}(x) =   \pt\psi_{(0.6,\;-1/\sqrt{2})}(x+x_0,t=0),\\\displaybreak[1]
   &\psi_3^{(1)}(x) = \pt\psi_{(-0.25,\;\sqrt{3}/2)}(x-x_0,t=0),\\\displaybreak[1]
   & Q^{(0)}(x)=  Q_{0,6}(x+x_0,t=0) +Q_{-0.25}(x-x_0,t=0),\label{3sc_initial_q}
\end{align}
again with a dislocation parameter $x_0>0$ to measure the separation
of two one-soliton waves at initial time level.  Some results from
numerical experiments are depicted in Figure \ref{fig_3sc_c} and
Figure \ref{fig_3sc_evol}, in which initially two soliton waves are
well separated and only a wave front moving towards the right
$x$-axis exists in $\psi_2$  while only a wave front moving towards
the left $x$-axis exists in $\psi_3$.  It shows that after the
collision a new wave front moving towards the left $x$-axis appears
in $\psi_2$ and a new wave front moving towards the right $x$-axis
appears in $\psi_3$.  Also, a conspicuous change in the amplitude of
waves is observed, see Figure \ref{fig_3sc_evol}.  Results for two
solitons not separated at initial time level are quite similar as in
one-coupled Klein--Gordon equations.

The results presented in this section indicate that the dynamics of
waves governed by coupled nonlinear Klein--Gordon equations is a
rather complicated issue. The results also demonstrate the
efficiency and high resolution of the proposed method for
numerically studying the coupled nonlinear Klein--Gordon equations.

\section{Conclusions}\label{s4}

A numerical scheme, which is based on the application of
pseudospectral approximations to spatial derivatives followed by a
trigonometric integrator to temporal discretization in phase space,
was proposed for solving coupled nonlinear Klein--Gordon equations.
The soliton solutions arising from one-coupled Klein--Gordon
equations were examined to test the accuracy and stability. Also,
application results of studying soliton-soliton collisions in one-
and three-coupled Klein--Gordon equations were reported as examples
to demonstrate the efficiency and high resolution of the scheme.

\section*{Acknowledgements}
This work was supported by Academic Research Fund of Ministry of
Education of Singapore grant R-146-000-120-112. Part of this work
was done when the author was visiting the Isaac Newton Institute for
Mathematical Sciences in Cambridge.  The visit was supported by the
Isaac Newton Institute.


\bigskip


\begin{thebibliography}{10}

\bibitem{AC_book}
{M.J. Ablowitz and P.A. Clarkson}, {Solitons, nonlinear evolution
equations and inverse scattering transform}, Cambridge University
Press, Cambridge, 1990.

\bibitem{AS_SIAM}
{M.J. Ablowitz and H. Segur}, {Solitons and inverse scattering
transformation}, SIAM, Philadelphia, 1981.

\bibitem{ACN_ckg}
{T. Alagesan, Y. Chung and K. Nakkeeran}, {Soliton solutions of
coupled nonlinear Klein--Gordon equations}, Chaos Solitions Fract.
21 (2004) 879--882.

\bibitem{ACN_integrable}
{T. Alagesan, Y. Chung and K. Nakkeeran}, {Painlev\'{e} analysis of
$N$-coupled nonlinear Klein--Gordon equations}, J. Phys. Soc. Jpn.
72 (2003) 1818.

\bibitem{BD2011}
{W. Bao and X. Dong}, {Analysis and comparison of numerical methods
for Klein--Gordon equation in nonrelativistic limit regime}, Numer.
Math. (in press) DOI 10.1007/s00211-011-0411-2.

\bibitem{CaoGuo}
{W. Cao and B. Guo}, {Fourier collocation method for solving
nonlinear Klein--Gordon equation}, J. Comput. Phys. 108 (1993)
296--305.


\bibitem{Deeba}
{E.Y. Deeba and S.A. Khuri}, {A decomposition method for solving the
nonlinear Klein--Gordon equation}, J. Comput. Phys. 124 (1996)
442--448.


\bibitem{Deu79}
{P. Deuflhard}, {A study of extrapolation methods based on multistep
schemes without parasitic solutions}, Z. Angew. Math. Phys. 30
(1979) 177--189.

\bibitem{Duncan}
{D.B. Duncan}, {Symplectic finite difference approximations of the
nonlinear Klein--Gordon equation}, SIAM J. Numer. Anal. 34 (1997)
 1742--1760.

\bibitem{Gau61}
{W. Gautschi}, {Numerical integration of ordinary differential
equations based on trigonometric polynomials}, Numer. Math. 3
(1961) 381--397.

\bibitem{HO_ckg}
{H. Hirota and Y. Ohta}, {Hierarchies of coupled soliton equations.
I}, J. Phys. Soc. Jpn. 60 (1991) 798--809.

\bibitem{Hirota_bilinear}
{R. Hirota}, {Direct method of finding exact solutions of nonlinear
evolution equations}, Lect. Notes Math., vol. 515, pp. 40--68,
Springer, Berlin-Heidelberg-New York, 1976.

\bibitem{LiVu}
{S. Li and L. Vu-Quoc}, {Finite difference calculus invariant
structure of a class of algorithms for the nonlinear Klein--Gordon
equation}, SIAM J. Numer. Anal. 32 (1995) 1839--1875.

\bibitem{Pascual}
{P.J. Pascual, S. Jim\'{e}nez and L. V\'{a}zquez}, {Numerical
simulations of a nonlinear Klein--Gordon model. Applications}, Lect.
Notes Phys., vol. 448, pp. 211--270, Springer, Berlin, 1995.


\bibitem{PA_P-analysis}
{K. Porsezian and T. Alagesan}, {Painlev\'{a} analysis and complete
integrability of coupled Klein--Gordon equations}, Phys. Lett. A 198
(1995) 378--382.

\bibitem{ST}
{J. Shen and T. Tang}, {Spectral and High-Order Methods with
Applications}, Science Press, Beijing, 2006.



\end{thebibliography}
\end{document}